\documentclass[12pt,preprint2]{emulateapj}
\usepackage{pslatex}
\usepackage[T1]{fontenc}
\usepackage[latin1]{inputenc}
\usepackage{subfigure}
\usepackage{amsmath}
\usepackage{graphicx,graphics}
\usepackage{amssymb}
\usepackage[english]{babel}

\newcommand{\be}{\begin{equation}}
\newcommand{\ee}{\end{equation}}
\newcommand{\beqn}{\begin{eqnarray}}
\newcommand{\eeqn}{\end{eqnarray}}
\newcommand{\lap}{\lesssim}
\newcommand{\gap}{\gtrsim}

\newcommand{\msun}{M_\odot}

\newcommand{\beq}{\begin{equation}}
\newcommand{\eeq}{\end{equation}}

\def\gap{\;\rlap{\lower 2.5pt
 \hbox{$\sim$}}\raise 1.5pt\hbox{$>$}\;}
\def\lap{\;\rlap{\lower 2.5pt
   \hbox{$\sim$}}\raise 1.5pt\hbox{$<$}\;}

\shorttitle{Universal Density Profile}
\shortauthors{Merritt et al.}
\begin{document}

\title{A Universal Density Profile for Dark and Luminous Matter?}

\author{David Merritt\altaffilmark{1},
Julio F. Navarro\altaffilmark{2,3,$\dagger$}, Aaron Ludlow\altaffilmark{2} 
and Adrian Jenkins\altaffilmark{4}
}
\altaffiltext{1}
{Department of Physics, Rochester Institute of Technology, 
Rochester, NY 14623, USA}
\altaffiltext{2}
{Department of Physics and Astronomy, University of Victoria, 
Victoria, BC, V8P 1A1, Canada}
\altaffiltext{3}
{Max-Planck Institute for Astrophysics, Karl-Schwarzschild Strasse 1, 
Garching, D-85740, Germany}
\altaffiltext{$\dagger$}
{Fellow of CIAR and of the Guggenheim Foundation}
\altaffiltext{4}
{Institute for Computational Cosmology, Durham University, Durham DH1 3LE,
  England}

\begin{abstract}
We explore similarities in the luminosity distribution of early type galaxies
and the mass profiles of $\Lambda$CDM halos. The spatial structure of these
systems may be accurately described by a simple law where the logarithmic
slope of the projected density is a power law of radius; the S\'ersic law. We
show that this law provides a significantly better fit than a three-parameter
generalization of the NFW profile and derive the best-fitting S\'ersic
parameters for a set of high-resolution $\Lambda$CDM halos spanning a wide
range in mass. The mean S\'ersic $n$ values are $3.0$ for dwarf- and
galaxy-sized halos and $2.4$ for cluster-sized halos, similar to 
the values that characterize luminous elliptical galaxies.  
We discuss possible reasons why the same
law should describe dark and luminous systems that span a range of over seven
decades in mass.
\end{abstract}

\section{Introduction}

The \cite{sersic68} law,
\begin{equation}
\ln (\Sigma/\Sigma_e) = - b\, (X^{1/n} - 1),
\label{eq:sersic}
\end{equation}
relating the 2D (projected or surface) density, $\Sigma$, and the dimensionless
radius, $X=R/R_e$, is often fit to the luminosity profiles of elliptical
galaxies and to the bulges of disk galaxies. The parameters of
the fit include the S\'ersic index, $n$, as well as the constant, $b$, which
is normally chosen so that $R_e$ is the radius containing one-half of the
projected light; $b=b(n)\approx 2n-0.324$ \citep{ciotti99}.

In a recent series of papers, A. Graham and co-workers have shown that the
S\'ersic law provides a remarkably good fit to the luminosity profiles of
stellar spheroids, from dE galaxies to the most luminous ellipticals
\citep{graham01,graham02,gguzman03,graham03,trujillo04}. The fits apply over
2-3 decades in radius, and often extend down to the innermost resolvable
radius.  Deviations from the best-fitting S\'ersic law are typically of order
$0.05$ magnitudes rms.  The S\'ersic index $n$ is found to correlate well with
galaxy absolute magnitude,
\begin{equation}
\log_{10}n \approx -0.106 \, M_B - 1.52
\label{eq:nvsm}
\end{equation}
\citep{gguzman03}, and also with other structural parameters like $R_e$ and
$\Sigma_e$
\citep{caon93,gguzman03}. Setting $n=4$ gives the \cite{devauc48}
law, which is a good fit to luminous elliptical galaxies, and $n=1$ is the
exponential law, which reproduces well the luminosity profiles of dwarf
ellipticals.

There are some known limitations to the applicability of 
equation (\ref{eq:sersic}) to the very central regions of some galaxies. 
In particular, S\'ersic's law
fails to represent adequately the very central profiles of elliptical galaxies
with cores; the pointlike nuclei of some dE galaxies; and the steep power-law
density cusps observed in the inner few parsecs of nearby galaxies like M32
and the bulge of the Milky Way.  The origin of these features is not well
understood, but it is likely that they are the result of dynamical processes,
possibly involving single or multiple black holes, which act to modify the
pre-existing S\'ersic profile in the innermost regions
\citep{marel99,mm02,ravin02,graham04,merritt04,preto04}.

The density profiles of the {\it dark matter} halos formed in $N$-body
simulations of hierarchical clustering have traditionally been fit to a rather
different class of functions, essentially broken power laws
\citep{nfw96,nfw97,moore99}.  However, the most recent simulations
\citep{power03,reed04} suggest that halo density profiles are better
represented by a function with a continuously-varying slope.  \cite{paper3}
proposed the fitting function
\begin{equation}
{d\ln\rho/d\ln r}= -2\left({r/ r_{-2}}\right)^\alpha
\label{eq:drhodr}
\end{equation}
where $r_{-2}$ is the radius at which the logarithmic slope
of the {\it space} density is $-2$, 
and $\alpha$ is a parameter describing the degree of variation
of the slope.
The corresponding density profile is
\begin{equation}
\ln(\rho/\rho_{-2})= -{(2/\alpha)}\, (x^\alpha - 1)
\label{eq:alpha}
\end{equation}
with $x\equiv r/r_{-2}$.  Remarkably, this is precisely the same functional
form as equation (\ref{eq:sersic}) -- with the difference that Navarro et
al. fit equation (\ref{eq:alpha}) to the {\it space} density 
of dark matter halos, while
equation (\ref{eq:sersic}) applies to the {\it projected} densities of
galaxies.

Nevertheless the connection is intriguing and a number of questions spring to
mind.  Does the S\'ersic profile fit the surface density profiles of dark
matter halos as well as it fits galaxies?  We will show here (\S3) that the
answer is ``yes'': the same fitting function provides an equally good
description of the projected densities of both dark and luminous spheroids.
In \S4 we ask whether it is most appropriate to fit the S\'ersic law to the
space or projected densities of dark matter halos, and whether these functions
are better fits than other three-parameter functions.  \S5 contains some
speculations about why a single density law should describe dark and luminous
systems over such a wide range in mass.

\section{Method}

We constructed nonparametric estimates of the space and projected density
profiles of the 19 $\Lambda$CDM halo models in \cite{paper3}, and compared
them with a number of fitting functions, including the S\'ersic law, equation
(\ref{eq:sersic}); the deprojected S\'ersic law $\Sigma_d(r)$, defined as the
spherical density law whose spatial projection is $\Sigma(R)$; 
and a generalized,
three-parameter NFW (1996, 1997) profile, which may be expressed as
\begin{equation}
%\ln\rho = \ln\rho_s - \gamma\ln x - (3-\gamma)\ln (1+x)
d\ln{\rho}/d\ln{x} = -(\gamma+3x)/(1+x),
\label{eq:nfw}
\end{equation}
with $x=r/r_s$.  The NFW profile has $\gamma=1$ and $r_s=r_{-2}$; the
\cite{moore99} profile has a similar functional form with inner slope
$\gamma=1.5$.

Details of the numerical simulations are given in \cite{paper3}.  Four halos
are ``dwarf'' sized ($M\approx 10^{10}\msun$), seven are ``galaxy'' sized
($M\approx 10^{12}\msun$), and eight are ``cluster'' sized ($M\approx
10^{15}\msun$).  We adopt the notation of that paper ($D$= dwarf, $G$=galaxy,
$C$=cluster) in what follows.  

\begin{figure}
\includegraphics[angle=-90.,scale=0.6]{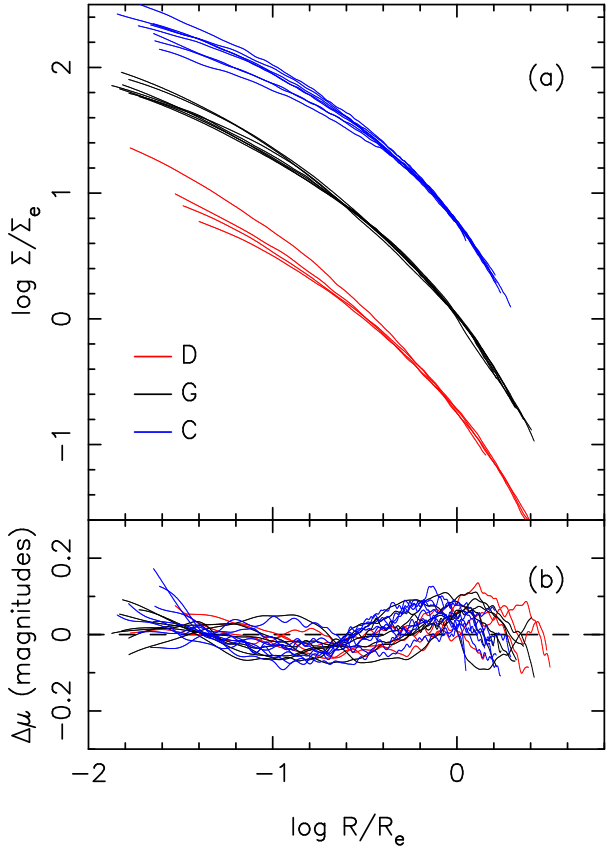}
\caption{(a) Nonparametric estimates of the surface density
profiles of the 19 halo models. 
Profiles of the $D$ ($C$) models have been shifted downward (upward)
by $0.75$ in the logarithm.
(b) Deviations of the best-fitting S\'ersic
model from $\hat\Sigma(R)$. 
Fitting parameters are given in Table 1.
}
\end{figure}

Nonparametric estimates of the space and projected density profiles,
$\hat\rho(r)$ and $\hat\Sigma(R)$, were constructed using the spherically
symmetrized kernels defined by \cite{mt94} (see e.g. \cite{reed04}, Appendix
A).  Each $N$-body point was replaced by a kernel of the form
\begin{eqnarray}
K_{\rho}(r,r_i,h_i)&=&{1\over 2(2\pi)^{3/2}}\left({rr_i\over h_i^2}\right)^{-1}
e^{-\left(r_i^2+r^2\right)/2h_i^2} \sinh\left(rr_i/h_i^2\right), \\
K_{\Sigma}(R,R_i,h_i)&=&{1\over 2\pi}e^{-\left(R_i^2+R^2\right)}I_0(RR_i/h_i^2)
\end{eqnarray}
with $h_i$ the width of the kernel associated with the $i$th particle and
$I_0$ the modified Bessel function.  The projected radii $R_i$ were obtained
from the $N$-body radii $r_i$ by assigning each particle a random position on
the sphere of radius $r_i$.  Density estimates were computed on a grid of 100
radial points spaced logarithmically from $r_{conv}$ to $r_{200}$
(these radii are defined below).  
We followed standard practice \citep{silverman86} and first computed a pilot
estimate of the density via a nearest-neighbor scheme, then allowed the $h_i$
to vary as a power $\delta$ of this pilot density.

When fitting one of the parametric functions defined above to $\hat\rho$ or
$\hat\Sigma$, we computed the density estimates on a grid in $(h_0,\delta)$
($h_0$ is the geometric mean of the $h_i$) to see which choice of kernel
parameters minimized the residual of the fit; typically there was a broad
range of $(h_0,\delta)$ values over which the best-fit parameters and their
residuals were nearly constant.  The residual was defined as the rms over the
radial grid of $\log(\hat\rho_j/\rho(r_j))$ with $\hat\rho_j$ the density
estimate at grid point $r_j$ and $\rho$ the parametric fitting function; this
is identical to how most observers define the residual.  Below we state the
rms deviation between the ``measured'' profile and the best-fitting parametric
model in terms of magnitudes, denoted by $\Delta\mu$.

We followed the practice in \cite{paper3} of only constructing 
density estimates
in the radial range $r_{conv}\le r\le r_{200}$, where $r_{conv}$ is the radius
beyond which the halo mass distribution is considered robust to errors or
approximations associated with the simulations (particle softening, relaxation
etc.) and $r_{200}$ is the virial radius, i.e. the radius within which the
mean density contrast is $200$ times the critical density. Table 2 of
\cite{paper3} gives values of $r_{200}$ and $r_{conv}$ for all halo models.

\begin{deluxetable}{ccc|cccccc}
\tablewidth{0pt}
\tablecaption{Model fits to the halo density profiles.}
\tablehead{
\colhead{Halo} & \colhead{$\ \ \ \ \ \ \Sigma$}& & & & 
\colhead{$\ \ \ \ \rho$} 
& & & \\
 & \colhead{$n$} & \colhead{$\Delta\mu$}
 & \colhead{$n_d$} & \colhead{$\Delta\mu$} & \colhead{$n$}  
 & \colhead{$\Delta\mu$} & \colhead{$\gamma$} & \colhead{$\Delta\mu$}
}
\startdata
D1 & 3.04 & 0.043 & 3.47 & 0.047 & 5.58 & 0.054 & 1.34 & 0.071 \\
D2 & 2.63 & 0.043 & 2.89 & 0.024 & 4.47 & 0.029 & 0.89 & 0.088 \\
D3 & 3.91 & 0.018 & 4.19 & 0.041 & 6.94 & 0.039 & 1.51 & 0.041 \\
D4 & 2.84 & 0.067 & 3.33 & 0.059 & 5.26 & 0.065 & 1.23 & 0.090 \\
G1 & 2.94 & 0.030 & 3.17 & 0.036 & 5.38 & 0.038 & 1.23 & 0.047 \\
G2 & 3.21 & 0.055 & 3.47 & 0.056 & 5.63 & 0.052 & 1.25 & 0.072 \\ 
G3 & 2.87 & 0.050 & 3.44 & 0.042 & 5.98 & 0.049 & 1.33 & 0.047 \\
G4 & 3.30 & 0.040 & 3.70 & 0.022 & 6.13 & 0.015 & 1.36 & 0.027 \\
G5 & 2.95 & 0.047 & 3.03 & 0.077 & 4.91 & 0.064 & 1.03 & 0.054 \\
G6 & 2.93 & 0.063 & 3.57 & 0.059 & 6.10 & 0.069 & 1.39 & 0.064 \\ 
G7 & 2.82 & 0.051 & 3.09 & 0.087 & 5.04 & 0.097 & 1.22 & 0.110 \\
C1 & 2.49 & 0.062 & 3.36 & 0.048 & 6.36 & 0.046 & 1.26 & 0.059 \\
C2 & 2.41 & 0.028 & 2.68 & 0.047 & 4.65 & 0.040 & 1.06 & 0.048 \\
C3 & 2.50 & 0.031 & 2.92 & 0.026 & 5.02 & 0.031 & 1.18 & 0.029 \\
C4 & 2.19 & 0.064 & 3.11 & 0.101 & 5.72 & 0.111 & 1.29 & 0.091 \\
C5 & 2.53 & 0.041 & 2.61 & 0.092 & 4.33 & 0.077 & 0.91 & 0.066 \\
C6 & 2.16 & 0.044 & 2.62 & 0.050 & 4.49 & 0.065 & 1.10 & 0.064 \\
C7 & 2.79 & 0.047 & 3.99 & 0.040 & 7.44 & 0.042 & 1.41 & 0.038 \\
C8 & 1.99 & 0.069 & 2.62 & 0.083 & 4.67 & 0.095 & 1.13 & 0.087 \\
\enddata
\end{deluxetable}

\section{Dark Matter Halos as S\'ersic Models}

With few exceptions, modelling of the
luminosity profiles of galaxies is done in projected space.  We therefore
began by analyzing the surface density profiles of the dark halos.
Nonparametric estimates of $\Sigma(R)$ for the 19 halos are shown in Figure
1a, and Figure 1b plots the deviations from the best-fitting S\'ersic model,
equation (\ref{eq:sersic}); Table 1 gives the best-fitting $n$ and
$\Delta\mu$.  The mean S\'ersic index is $3.11\pm 0.49$ ($D$), $3.00\pm 0.17$
($G$), $2.38\pm 0.24$ ($C$), possibly indicating a (weak) trend toward
decreasing curvature (lower $n$) 
in the profiles of halos of increasing mass.  

The $\Delta\mu$ values average $0.043$ ($D$), $0.048$ ($G$), $0.048$ ($C$).
For comparison, \cite{caon93} find $\Delta\mu\approx 0.05$ in a sample of $45$
E and S0 galaxies, and \cite{trujillo04} find a mean $\Delta\mu$ of $0.09$ in
a sample of 12 elliptical galaxies without cores.  The radial range over which
luminous galaxies are fit varies from $\sim 1.5$ to $\sim 3.5$ decades,
comparable on average with the $\sim 2$ decades characterizing our dark-matter
halos.  While the noise properties are different for the two types of data,
the particle numbers in our halo models are small enough ($\sim 10^6$) to
contribute nonnegligibly to $\Delta\mu$.  We conclude that the S\'ersic law
fits dark halos as well as, and possibly even better than, it fits luminous
galaxies.

The residuals in Figure 1b appear to show some structure.
We will return in a future paper to the question of
whether a modification of the S\'ersic law might
reduce these residuals even further.

\begin{figure}
\includegraphics[angle=0.,scale=0.6]{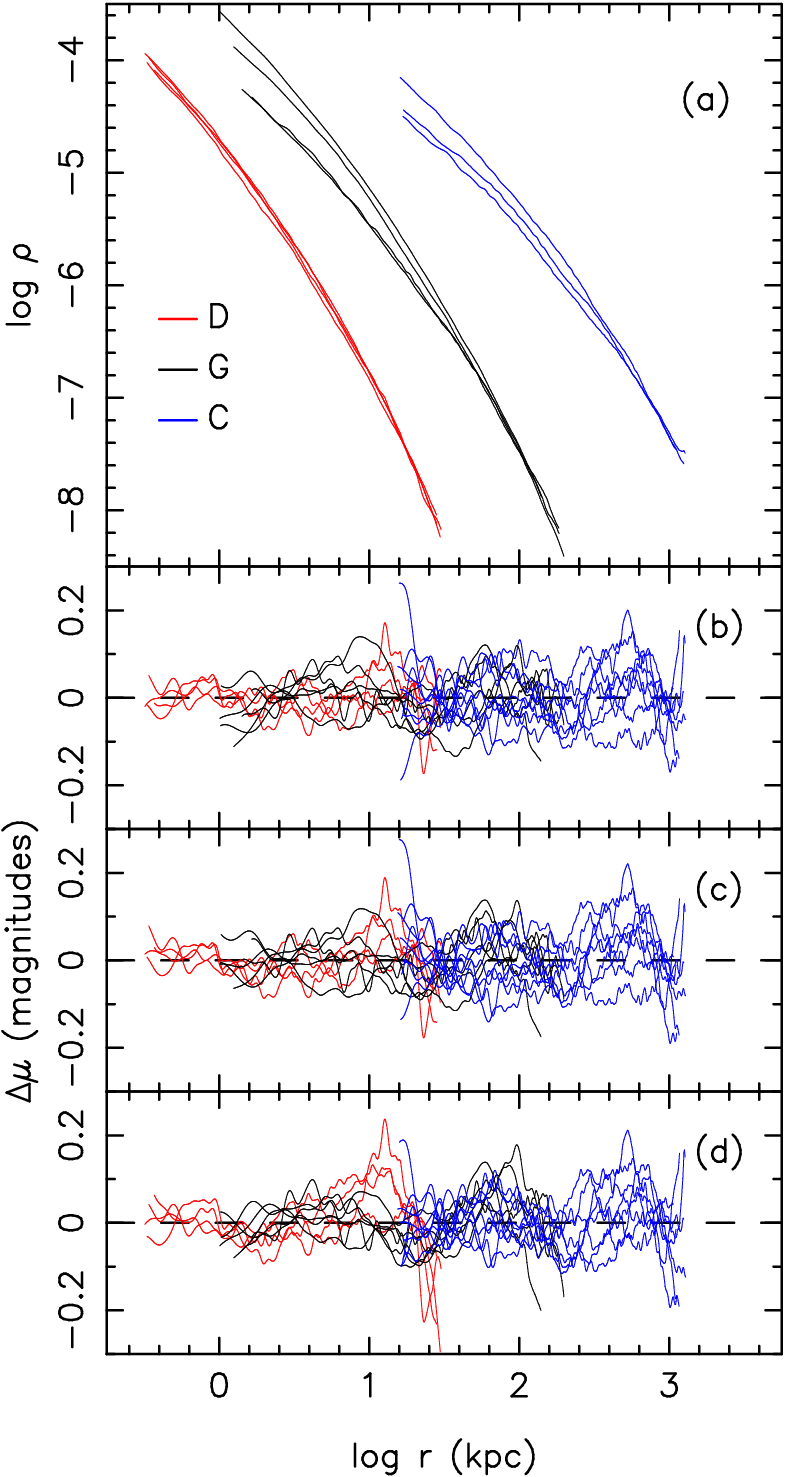}
\caption{(a) Nonparametric estimates of the space density
of the 19 dark halos. 
Vertical normalization is arbitrary.
(b-d) Deviations in magnitudes of three parametric models 
from $\hat\rho(r)$: (b) deprojected S\'ersic model;
(c) equation (\ref{eq:alpha});
(d) generalized NFW model, equation (\ref{eq:nfw}). 
Best-fit parameters are given in Table 1.
}
\end{figure}

\section{Which Function Fits the Space Density Best?}

\cite{paper3} showed that equation (\ref{eq:alpha}) provides a good fit to the
{\it spatial} density profiles of dark halos.  We showed above (\S3) that the
S\'ersic law (\ref{eq:sersic}) is a good fit to the {\it surface}
density profiles of dark halos.  An obvious inference is that a {\it
deprojected} S\'ersic law should provide a good fit to the space density.
Here we ask which function -- equation (\ref{eq:alpha}), 
or a deprojected S\'ersic law -- gives a better fit to $\rho(r)$.  
We also consider the quality of fit of
another three-parameter function, the generalized NFW profile presented in
equation (\ref{eq:nfw}).  

When fitting deprojected S\'ersic profiles to the dark halos, we define $n_d$
to be the S\'ersic index of the projected function; hence $n_d$ should be
close to the index $n$ derived when fitting a S\'ersic law to the surface
density (and the two would be equal if the halo's surface density were
precisely described by S\'ersic's law).  When reporting fits to $\rho(r)$ with
equation (\ref{eq:alpha}) we define $n\equiv \alpha^{-1}$, with $\alpha$ the
shape parameter of equation (\ref{eq:drhodr}).

The results are shown in Figure 2 and Table 1.  Mean values of $\Delta\mu$ for
the three fitting functions (deprojected S\'ersic, eq. (\ref{eq:alpha}), 
generalized NFW)
are ($0.043,0.047,0.073$) for the dwarf halos, ($0.054,0.055,0.060$) for the
galaxy halos, and ($0.061,0.063,0.060$) for the cluster halos.  Thus, the
two S\'ersic functions are almost indistinguishable in terms of their goodness
of fit: at least over the radial range available, a deprojected S\'ersic
profile with index $2.5\lap n_d\lap 3.5$ can be well approximated by a
S\'ersic profile with $n$ in the range $4.5\lap n \lap 7.5$.  Both functions
provide a significantly better fit to $\rho(r)$ than the generalized NFW
profile in the case of the dwarf halos, and the two S\'ersic functions perform
at least slightly better than NFW for the galaxy halos.  No single function is
preferred when fitting $\rho(r)$ for the cluster halos. 
\footnote[1]{Also of interest are the mean $\gamma$-values in the fits
to the generalized NFW profile.
We find $\langle\gamma\rangle=(1.24,1.26,1.17)$ for dwarf,
galaxy and cluster halos respectively.
We note that these are significantly shallower than the
steep inner slope ($\gamma=1.5$) proposed by Moore et al. (1999)
and, as discussed by Navarro et al. (2004), are best interpreted
as upper limits to the inner aymptotic behavior of the profile.}

\begin{figure}
\includegraphics[scale=0.6]{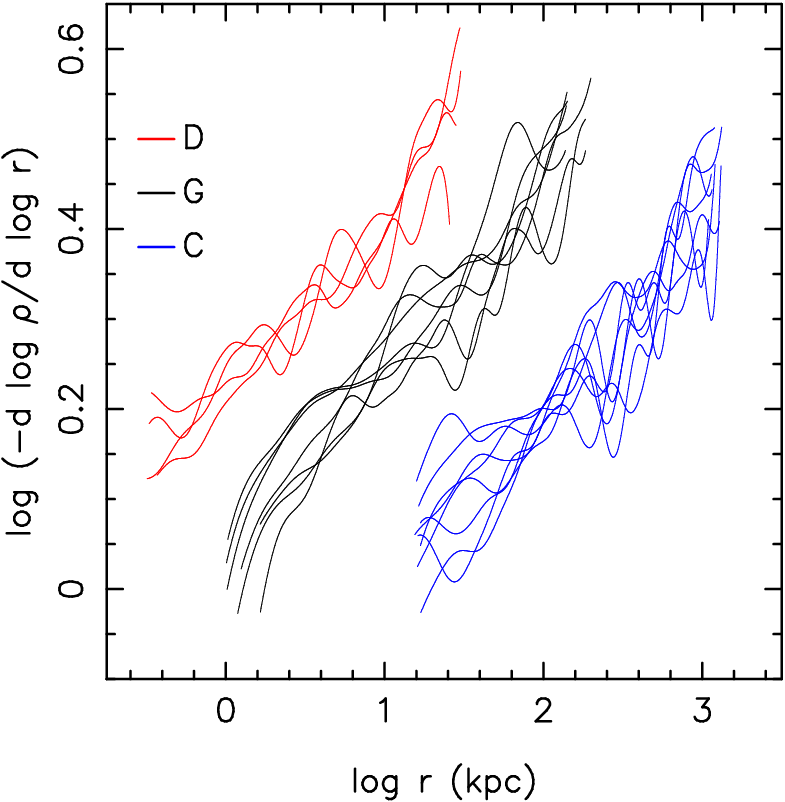}
\caption{Nonparametric estimates of the logarithmic
derivative of the space density for the 19 halo models. 
}
\end{figure}

Another way to compare the halo density profiles with S\'ersic's law is via
the radial dependence of the slope.  Figure 3 shows nonparametric estimates of
the logarithmic slope, $d\log\rho/d\log r$, for the dark halos; slopes were
computed via direct differentiation of the kernel density estimates, using a
larger kernel width to compensate for the greater noise generated by the
differentiation.  Equation (\ref{eq:drhodr}) 
predicts a straight line on this plot.  
That is a reasonable description of Figure 3.  
The value of $d\log\rho/d\log r$ in the $G$ and $C$
halos reaches $\sim -1$ at the innermost radii, consistent with the asymptotic
power-law inner behavior of an NFW profile.  No obvious convergence to a
power law (constant logarithmic slope)
is seen in Figure 3, and it is likely that simulations of improved
resolultion may lead to even shallower slopes at smaller radii, as pointed out
by \cite{paper3}.

\begin{figure}
\includegraphics[scale=0.6]{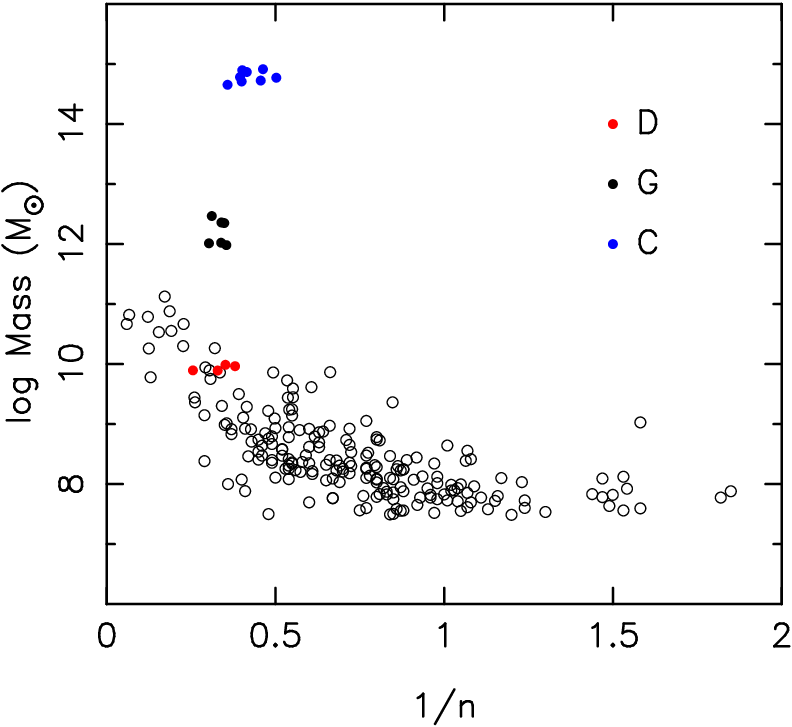}
\caption{S\'ersic index (derived from fits to the surface density)
versus mass for galaxies (open circles) and dark halos.
Galaxy points are taken from \cite{bj98,stiavelli01,
gguzman03,caon93,dono94}.
Halo masses are $M_{200}$ from \cite{paper3}.
Galaxy masses were computed from total luminosities
assuming the Magorrian et al. (1998)
mass-to-light ratio, with $H_0=70$ km s$^{-1}$ Mpc$^{-1}$.
}
\end{figure}

\section{What Does it Mean?}

Figure 4 shows S\'ersic's $n$ (derived from fits to the 
surface density) as a function of mass for our dark halos
and for a sample of early-type  galaxies.
There is overlap at $M\approx 10^{10}\msun$, 
the mass characteristic of ``dwarf'' halos
and giant ellipticals.
However the galaxies exihibit a much wider range
of $n$ values,
extending to $n<0.5$ in the case of dwarf ellipticals.
A natural interpretation is that $n$ is determined by 
the degree to which (dissipationless) merging has
dominated the evolution.
The nearly exponential ($n\approx 1$) profiles
of dE galaxies are similar to those of disk galaxies,
suggesting that dissipation played a critical
role in their formation.
Luminous ellipticals are the end products of
many mergers, the most recent of which
are likely to have been gas-poor,
and have de Vaucouleurs-like profiles ($n\approx 4$).
This view is supported by numerical
simulations (Scannapieco \& Tissera 2002;
Eliche-Moral et al. 2005) that show how
exponential profiles are converted into
de Vaucouleurs-like profiles via repeated
mergers.

A thornier question is: Why should a law
like S\'ersic's fit dark or luminous spheroids
in the first place?
S\'ersic's law with $2\lap n\lap 4$
has an energy distribution that
is roughly Boltzmann, $N(E)dE\sim e^{\beta E}dE$,
and it is sometimes loosely argued that this
``maximum-entropy'' state is a result of the
mixing that accompanies violent relaxation or 
merging \citep{binney82,merritt89,ciotti91}.
With regard to dark halos, 
Taylor \& Navarro (2001) have shown that
the dependence of phase-space density on radius
is well approximated by a power law whose corresponding 
inner density profile has the shallowest slope.
This can again be interpreted as an indication
that the halos are well mixed.
While our study does not shed a great deal of
light on this question, it does suggest
that the scale-free property of S\'ersic's
law, $d\ln\rho/d\ln r \propto r^\alpha$,
is the feature that links dark and luminous 
spheroids and that this property may be a 
hallmark of systems that form via gravitational 
clustering.

We have shown that the fitting function that
best describes luminous galaxies, the S\'ersic
law, is an equally good fit to dark halos.
We have not shown that the S\'ersic law
is a good fit in an {\it absolute} sense to 
either sort of system.
But given that dark and luminous density profiles
are not pure power laws, a three-parameter law
like S\'ersic's is as parsimonious a description
as one can reasonably expect.
Future work should explore whether other, 
three-parameter fitting functions can describe
dark and/or luminous systems better than
S\'ersic's law.

We thank Alister Graham for making available the 
galaxy data in Figure 4 and for informative discussions.
DM was supported by grants
AST-0206031, AST-0420920 and AST-0437519 from the 
NSF and grant NNG04GJ48G from NASA.
JFN acknowledges support from the Alexander von Humboldt Foundation.

\end{document}